\documentstyle[pre,aps,epsf,fancyhdr]{revtex}
\begin{document}

\draft

%\preprint{IPT-2000-}

\title
{Exact solution of a class of one-dimensional nonequilibrium stochastic models\footnote{ to appear in Physical Review E}}

\author
{M.Mobilia and P.-A. Bares  }

\address
{Institute of Theoretical Physics, Swiss Federal Institute of Technology of  Lausanne, CH-1015 Lausanne  EPFL, Switzerland }

\date{\today} 

\maketitle

%%%start of the main document%%%%%%%%%%%%%%%%%%%%%

\begin{abstract}
We consider various one-dimensional non-equilibrium
models, namely the {\it diffusion-limited pair-annihilation and creation model} (DPAC) and its unbiased version (the Lushnikov's  model), the DPAC model with particle injection (DPACI), as well as (biased) diffusion-limited coagulation model (DC). We study the DPAC model using an  approach based on a duality transformation and the generating function of the dual model. We are  able to compute exactly the density and correlation functions in the general case with arbitrary initial states.
Further, we assume  that a source injects particles in the system. Solving, via the duality transformation, the equations of motions of the density and the non-instantaneous two-point correlation functions, we see how the
 source affects the dynamics.
Finally we extend the previous results to the DC model with help of a {\it similarity transformation}.
   
\end{abstract}

\section{Introduction}

Stochastic reaction-diffusion models play an important role in the description
of  classical interacting many-particle nonequilibrium systems in physics and in interdisciplinary areas (chemistry, biology, economics, ...). Usually physical systems are much too complex to be treatable  analytically or even numerically. However, in the context of critical phenomena  simple toy models have been shown to be useful in determining  universal properties and understanding possible relationships between microscopically different processes (see e.g. \cite{Privman} and references therein).

In  this work we study, on a periodic chain, models  which are prototypes of one-dimensional diffusion-limited reactions. Our main purpose is to present a novel approach (based on a duality transformation \cite{Santos0} combined with generating function techniques \cite{Aliev})  to analyze the DPAC and related models. In this paper we illustrate this approach by recovering known results and deriving, in a simple and systematic manner, some new ones and show that the method employed here is {\it complementary} to the {\it traditional} (see e.g. \cite{Schutzrev,Grynberg1,Racz,Family} and references therein)  ones. We also solve the dynamics of the {\it dual} of the DPAC model which is a biased generalization of {\it Glauber's model}. We believe that the method used here is particularly suited to {\it site-depend} and/or {\it disordered } systems. A more detailed presentation of the method, as well as other applications, will be given elsewhere \cite{mobar}.

Diffusion-annihilation and coagulation models (in their free-fermionic version) properly describe the kinetics of excitons in several materials: the dynamics of photoexcited solitons and polarons in, e.g., chains of $[Pt(en)_{2}][Pt(en)_{2} Cl_{2}](BF_{4})_{4}$, called $(MX)$, where $(en)$ denotes {\it ethylenedyamine}, or the fusion of photogenerated excitons in chains of tetramethylammonium manganese trichloride (TMMC) \cite{Kuroda}. Models of pair-annihilation and creation  are useful to decribe problems of dimer adsorption and desorption \cite{Grynberg1}.

The motivation for studying such one-dimensional systems is not only their experimental relevance, but also their theoretical importance in the understanding of the fluctuations of low-dimensional systems. In both one- and two- dimensional systems, diffusive mixing is inefficient and leads to the building of large-scale correlations. The mean-field is not adequate and in this sense exact results are desirable.

The paper is organized as follows: In the sequel of this section we recall the {\it stochastic Hamiltonian} formulation of Markovian processes obeying a  master equation. In section II we introduce two models: the {\it diffusion-limited  pair-annihilation-creation} (DPAC) model and the {\it diffusion-limited pair-annihilation-creation} model connected with a {\it source} (DPACI model). In section III we map through a {\it duality transformation} the  DPAC and DPACI models on other stochastic models. The section IV and V are devoted to the detailed study of the DPAC model. In section IV, we evaluate the exact generating function of the dual DPAC model. Correlation functions of the DPAC model are studied in section V. In section VI, we study the DPACI model via the duality transformation. In section VII we show how to extend the solution obtained for the {\it diffusion-limited pair-annihilation} (DPA) model to  the {\it diffusion-limited} (DC) model. The section VIII is devoted to the conclusion.

In this work we study some one-dimensional two-states nonequilibrium systems for which the dynamics takes place on a periodic chain with $L$ (even) sites. The dynamics of particles (of a single species) is governed by a Master equation.
Each site of the lattice can be either empty or occupied by a particle at most (the hard-core constraint), say, of species $A$ . It is known that such nonequilibrium problems can be reformulated as field-theoretical many-body problems. Below we briefly recall the basis of the field-theoretical formulation of the master equation.

The state of the system is represented by the ket $|P(t)\rangle=\sum_{\{n\}}P(\{n\},t)|n \rangle$, 
where the sum runs over the $2^L$ configurations.
At site $i$ the local state is specified by the ket 
$|n_i\rangle =(1 \; 0 )^{T}$ if the site 
$i$ is empty,  $|n_i\rangle =(0 \; 1 )^{T}$ if the site $i$ is occupied by 
a particle of type $A$. We define the {\it left vacuum} $\langle \widetilde \chi|$, where
$\langle \widetilde \chi|\equiv \sum_{\{n\}} \langle n |$. The  master equation can be rewritten formally as an imaginary-time Schr\"odinger equation: $\frac{\partial}{\partial t}|P(t)\rangle = -H |P(t)\rangle$,
where $H$ is the {\it Stochastic Hamiltonian} which governs the dynamics of the system . In general, it is neither hermitian nor normal. Its construction from the master equation is a well established procedure (see e.g. \cite{Schutzrev} and references therein). Here we specifically focus on two-state systems so that  the {\it stochastic Hamiltonian} can be rewritten in term of Pauli's operator as a (pseudo-)spin chain (see below). Probability conservation (stochasticity of $H$)$ yields \langle \widetilde \chi|H =0 $.
\section{The DPAC and DPACI models}

In this section we present two models which we will consider in the sequel of the work.
We begin  with the {\it diffusion-limited pair-annihilation-creation} (DPAC) model.

We consider, without loss of generality, a periodic chain of $L$ (even) sites on which an even number of particles with hard-core repulsion  move according to a Master equation \footnote{Because the DPAC model preserves the parity of the number of particles, we should split the dynamics in a {\it even} and an {\it odd} sector. Here we focus specifically, and without loss of generality, on the {\it even sector}. The treatment of the {\it odd} sector is similar excepted for the boundary conditions (see e.g. \cite{Santos0}). 
The parity of the number of particles does not play an important role for the {\it dynamics} of the model. The restriction to the even sector is performed  projecting out the odd sector from the initial state, say $|P(0)\rangle$, with help of the projector $\frac{1}{2}\left(\openone + Q\right)$, with $Q=\prod_{m}\sigma_{m}^{z}$. In the sequel, for simplicity, we will often call $|P(0)\rangle$
 its even-sector projection, i.e. $|P(0)\rangle^{even}\equiv \frac{1}{2}(\openone + Q)|P(0)\rangle$.}.

In the DPAC model, at each time-step four events can occur:

A particle can jump from site $r$ to $r+1$ (provided the latter was previously empty) with a probability rate $h'$.

A particle can jump from site $r$ to $r-1$ (provided the latter was previously empty) with a probability rate $h$.

A pair of particles can be created at sites $r$ and $r+1$ (provided the latter were previously empty) with a probability rate $\epsilon'$.

A pair of particles can be annihilated at sites $r$ and $r+1$ (provided the latter were previously occupied) with a probability rate $\epsilon$.

The Master equation associated to these processes can be formulated  as an imaginary time Schr\"odinger equation.
Identifying a vacancy with a (pseudo-) spin up and a particle with a (pseudo-) spin down the {\it stochastic Hamiltonian} can be written in term of Pauli (pseudo-)spin$-1/2$ operators, as follows:
\begin{eqnarray}
\label{eq.1.1}
&&H^{DPAC}=\sum_{r=1}^{L} H_{r,r+1}^{DPAC} \nonumber\\
H_{r,r+1}^{DPAC}&=&\left(\frac{\epsilon'- \epsilon + (h-h') }{4}\right) \sigma_{r}^{z} +\left(\frac{\epsilon'- \epsilon - (h-h') }{4}\right) \sigma_{r+1}^{z}+  \left(\frac{\epsilon'+\epsilon + h +h' }{4}\right)(1-\sigma_{r}^{x}\sigma_{r+1}^{x}) \nonumber\\ &+&  \left(\frac{\epsilon'+\epsilon - h - h' }{4}\right)
(\sigma_{r}^{z} \sigma_{r+1}^{z} + \sigma_{r}^{y} \sigma_{r+1}^{y} )
+i\left(\frac{\epsilon'- \epsilon - (h-h') }{4}\right)\sigma_{r}^{x} \sigma_{r+1}^{y} + i\left(\frac{\epsilon'- \epsilon + (h-h') }{4}\right) \sigma_{r}^{y} \sigma_{r+1}^{x},
\end{eqnarray}
where $\sigma_{j}^{x},\sigma_{j}^{y}, \sigma_{j}^{z}$ are the usual Pauli's matrices acting on site $j$  and $2\sigma_{j}^{\pm}\equiv \sigma_{j}^{x}\pm i\sigma_{j}^{y}$.
The operator $n_{j}\equiv\frac{1}{2}\left(\openone-\sigma_{j}^{z}\right)$ is the local (at site $j$) density operator ($\openone$ denotes the identity operator).
This model (with a particle-hole transformation) was introduced by Grynberg et al. \cite{Grynberg1}, and corresponds to  a biased generalization of Lushnikov's model \cite{Lushnikov}

The constraint
\begin{eqnarray}
\label{eq.1.2}
\epsilon+\epsilon'= h+h',
\end{eqnarray}
is the {\it free-femionic condition}, for which the model becomes soluble \footnote{Recently we have proposed \cite{Mobar2} an approach to study analytically the DPAC model {\it beyond} the free-fermion case (\ref{eq.1.2}). Although the approach \cite{Mobar2} provides the correct long-time behaviour of the density and correlation fuctions (including the subdominant terms), it turns out to be less general than expected \cite{Mobar3}. It is not the purpose of this work to study the limits of the approach devised in \cite{Mobar2}.}.
In the sequel, we assume that the condition (\ref{eq.1.2}) is fulfilled. When, in (\ref{eq.1.1}), no pairs are created, i.e. $\epsilon'=0$, the model under consideration is called $DPA$ ({\it diffusion-limited pair-annihilation}) model.

Let us now consider the following {\it stochastic Hamiltonian} (with periodic boundary conditions):
\begin{eqnarray}
\label{eq.1.3}
-H^{input}=\gamma\sum_{r=1}^{L}\left[(\sigma_{r}^{+}+\sigma_{r}^{-})(\sigma_{L}^{+}+\sigma_{L}^{-})-1\right]=\gamma \sum_{r=1}^{L}\left[\sigma_{r}^{x}\sigma_{L}^{x} -1\right]
\end{eqnarray}

This  {\it stochatic Hamiltonian} term corresponds to  a  single particles ``source'' which injects in the system, with rate $\gamma$,  particles at site $j<L$ and $L$, whenever both sites were previously vacant. If both sites $j<L$ and $L$ were previously occupied, the ``source'' annihilates (with rate $\gamma$) both particles. When one of the sites $j<L$ or  $L$ is occupied and the other empty, the effect of $H^{input}$  (with rate $\gamma$) is to fill in the previously vacant site and to empty the previously occupied one.
In (\ref{eq.1.3}), the term in $\sigma_{L}^{x}$ is an artefact due to the periodicity of the problem (because of the duality transformation, see next section) and preserves the parity of the system  
described by the {\it stochastic Hamiltonian} $H^{DPACI}\equiv H^{DPAC}
+ H^{input} $. The steady-state  as well as  the density relaxation time  of such a system (in its translationally invariant version) have been studied \cite{Racz} for the case where there was input of particles (i.e. $\gamma >0$), but no pair-creation (i.e. $\epsilon'=0$), for an infinite system (where there is no problem of boundary conditions and one can consider simply $-H^{input}=\gamma \sum_{r}\left[\sigma_{r}^{x} -1\right]$). Here we obtain the exact  density of the  DPACI  
model and study the interplay between the pair-creation and the source (see (\ref{eq.7.14.1})). We also obtain the  two-point
 non-instantaneous correlation functions of the (unbiased) DPACI model in the absence as well in the presence of pair-creation.

\section{The duality transformation}

Following \cite{Santos0,Schutzrev}, we introduce a set of operators which forms a periodic Temperley-Lieb algebra (a quotient of a Hecke algebra). Further, 
 we define an unitary transformation which allows to map the DPAC model on another stochastic model for which we are able to  calculate the generating function.

Let us define the following operators in term of Pauli 
spin-matrices:
\begin{eqnarray}
\label{eq.3.1}
e_{2j-1}\equiv\frac{1}{2}(1+\sigma_{j}^{x}), \; 1\leq j\leq L \;\;;\;\;
e_{2j}\equiv\frac{1}{2}(1+\sigma_{j}^{z}\sigma_{j+1}^{z}), \;  1\leq j\leq
 L-1  \;\;;\;\;
e_{0}\equiv e_{2L}=\frac{1}{2}(1+\hat{C}\sigma_{L}^{z}\sigma_{1}^{z}),
\end{eqnarray}
where $\hat{C}\equiv \prod_{j=1\dots L}\sigma_{j}^{x}$.

We  define the (unitary) duality transformation $V$ by
\begin{eqnarray}
\label{eq.3.6}
V\equiv\exp\left(\frac{i\pi}{4} \sum_{j=1}^{L} \sigma_{j}^{y}\right)\left\{\prod_{k=1}^{2L-1}[(1+i)e_{k}-1]\right\}
\end{eqnarray}

The dual (unitary) transformation $V$ is also called {\it domain-wall transformation} \cite{Santos0,Schutzrev} because of its physical interpretation. In \cite{Santos0} this duality transformation has been  introduced to study the zero-temperature Ising model with help of its free-fermionic dual counterpart
 (the DPA model, where $\epsilon'=0$). The transformation (\ref{eq.3.6}) connects a stochastic model with periodic boundary conditions to another stochastic models with the same boundary conditions \cite{Santos0}.

Let us define  the dual version of the stochastic Hamiltonian (\ref{eq.1.1}): 
 \begin{eqnarray}
\label{eq.3.8}
\hat{H}_{r}^{DPAC}\equiv V^{-1} H_{r,r+1}^{DPAC}(\epsilon,\epsilon', h, h') V\;\;;\hat{H}^{DPAC}\equiv \sum_{r=1}^{L}\hat{H}_{r}^{DPAC}
\end{eqnarray}
 with periodic boundary conditions:
$\sigma^{\#}_{L+r}=\sigma^{\#}_{r}, (r<L)$

From now on, we work with the dual model (\ref{eq.3.8})
and obtain information on the original model (\ref{eq.1.1}) (with the constraint (\ref{eq.1.2})). 
Notice that  the dual model (\ref{eq.3.8}) is still a stochastic one because, by construction, $\langle \widetilde{\chi}|H^{DPAC}=\langle \widetilde{\chi}|H_{r}^{DPAC}=
\langle \widetilde{\chi}|\hat{H}^{DPAC}=\langle \widetilde{\chi}|\hat{H_{r}}^{DPAC}=0$, since $\langle \widetilde{\chi}|V\equiv\langle \hat{\widetilde{\chi}}|= \alpha \langle \widetilde{\chi}|\;$, where $\alpha =\frac{(-1)^{L-1}}{\sqrt{2}}e^{i\frac{\pi}{4}(L-1)} $ \cite{Santos0} is a constant which  plays no role in the following and therefore will be omitted.

Here we consider the domain-wall duality transformation (\ref{eq.3.6}) and, formally,  obtain (when (\ref{eq.1.2}) is fulfilled) a generating functions for the dual model. The latter will allow  to  solve the DPAC model
completely. In the third part of this work, we translate our results into the language of the  diffusion-limited coagulation model.

The duality transformation (\ref{eq.3.6}) maps the DPAC model (\ref{eq.1.1}) 
onto a biased generalization of the Glauber-Ising model \cite{Grynberg1}. In the absence of bias, model (\ref{eq.1.1}) reduces to Lushnikov's model and the the duality transformation  (\ref{eq.3.6}) maps this model onto Glauber-Ising model \cite{Racz,Family,Alcaraz,Glauber}.

Let us also consider the dual transformation of $H^{input}$ (\ref{eq.1.3}):
 \begin{eqnarray}
\label{eq.3.11}
\hat{H}^{input}&\equiv&\sum_{r=1}^{L} V^{-1} H_{r}^{input} V
= \gamma \sum_{r=1}^{L}\left(\openone - \sigma_{r}^{x}\sigma_{r+1}^{x} \dots \sigma_{L-1}^{x}\right),
\end{eqnarray}
with periodic boundary conditions, and thus $\hat{H}^{DPACI}=\hat{H}^{DPAC}+\hat{H}^{input} $.

\section{The generating function of (the dual of) the DPAC model}

In the first part of the paper we  study exact properties of DPAC model using
 generating function of the dual model (\ref{eq.3.8}), which are explicitly computed in this section. We recover known results and produce  new ones. 

We introduce Grassmann numbers $\eta_{m}$ with  their usual anticommuting properties: $\{\eta_{m},\eta_{n}\}=0, \; \forall m,n$

Following \cite{Aliev}, we consider the quantity:
 \begin{eqnarray}
\label{eq.4.1}
G^{\pm}(\{\eta,j\},t)\equiv \langle \hat{\widetilde \chi}|  \prod_{j\geq 1}\left(1\pm \eta_{j}\sigma_{j}^{z}\right)e^{-\hat{H}^{DPAC}t}|\hat{P}(0)\rangle=
\left\langle \prod_{j\geq 1}\left(1\pm \eta_{j}
\sigma_{j}^{z}\right)(t)\right\rangle,
\end{eqnarray}
where $|P(0)\rangle$ symbolically denotes the initial state in the original (\ref{eq.1.1}) model with an even number of particles (see footnote 1), and  $|\hat{P}(0)\rangle\equiv V^{-1}|P(0)\rangle  $. $G^{\pm}(\{\eta,j\},t)$ is the 
 {\it generating function} of the dual model and its derivatives provide correlation function of the dual model (\ref{eq.3.8}), e.g $\langle \sigma_{i}^{z} \sigma_{j}^{z} \rangle (t)=\frac{\partial^2 G^{\pm} (\{\eta,j\},t) }
{\partial \eta_{j}\partial \eta_{i} }|_{\{\eta\}=0}, (i<j)$.

It has to be emphasized that at this stage all the correlation functions  obtained from the generating function $G^{\pm}(\{\eta,j\},t)$
are correlators of the dual model (\ref{eq.3.8}).
For the sequel, we introduce the following notations:
\begin{eqnarray}
\label{eq.4.6}
b\equiv \frac{h+h'+\epsilon+\epsilon'}{2}\;;\;
c\equiv \frac{h-h'+\epsilon-\epsilon'}{2}\;;\;
d\equiv \frac{h'-h+\epsilon-\epsilon'}{2}\;;\;
\widetilde{c}&\equiv& c+d=\epsilon-\epsilon' \;;\; 
\widetilde{d}\equiv c-d=h-h'
\end{eqnarray}

It is useful to separate the generating function into two parts. The one that generates the correlators with an {\it even}  numbers operators $\sigma^{z}$
 denoted ${\cal V}^{+}(\{\eta,j\},t)$. The functional, called ${\cal V}^{-}(\{\eta,j\},t)$, generates correlators with an {\it odd} number of operators $\sigma^{z}$, i.e.,
\begin{eqnarray}
\label{eq.4.7}
{\cal V}^{\pm}(\{\eta,j\},t)\equiv\frac{1}{2}\left(G^{+}(\{\eta,j\},t)\pm
 G^{-}(\{\eta,j\},t) \right)
\end{eqnarray}
When the {\it free-fermionic} constraint (\ref{eq.1.2}) is fulfilled, the equation of motion of the generating function  ${\cal V}^{\pm}$ can be rewritten, using the properties of Grassmann algebra, as a first-order partial differential equation
 which can be solved by the {\it method of characteristics} \cite{Aliev,Courant}:
\begin{eqnarray}
\label{eq.4.17}
{\cal V}^{\pm}(\{\eta,j\},t)= \frac{1}{2}\left\langle \left( \prod_{j\geq 1}\left(1+ \eta_{j}^{0}
\sigma_{j}^{z}(t=0)\right)\pm \prod_{j\geq 1}\left(1- \eta_{j}^{0}
\sigma_{j}^{z}(t=0)\right)\right) \right\rangle  \exp\left[\sum_{k_{1}}\sum_{k_{1}>k_{2}} \eta_{k_{1}}\eta_{k_{2}} K_{k_{1},k_{2}}^{\pm}(t) \right],
\end{eqnarray}
where $\eta_{k}^{0}\equiv e^{-bt}\sum_{j=1}^{L}(F^{\pm}(t))^{-1}_{kj} \eta_{j}$ and (in the thermodynamics limit, $L\rightarrow \infty$),
\begin{eqnarray}
\label{eq.4.19}
 (F^{\pm})^{-1}_{j,k}(t) &=&\frac{1}{2\pi} \int_{0}^{2\pi} d\phi \exp\left(i(j-k)\phi+\left\{2c \cos \phi - \widetilde{d}e^{i\phi}\right\}t\right)
=\delta^{\frac{j-k}{2}}I_{j-k}(\triangle t)
\end{eqnarray}
\begin{eqnarray}
\label{eq.4.18}
K_{j_{1},j_{2}}^{\pm}(t)=z^{j_{2}-j_{1}} -e^{-2bt}\sum_{k_{2}>k_{1}} z^{k_{2}-k_{1}}\left(\
I_{j_{1}-k_{1}}(\widetilde{c}t)I_{j_{2}-k_{2}}(\widetilde{c}t) - I_{j_{1}-k_{2}}(\widetilde{c}t)I_{j_{2}-k_{1}}(\widetilde{c}t) \right), \; (j_{2}>j_{1}),
\end{eqnarray}
where  the $I_{\nu}(w)$'s are the usual modified Bessel function of first kind,  $z\equiv \frac{b-\sqrt{b^{2}-\widetilde{c}^{2}}}{\widetilde{c}}$, $\triangle \equiv\sqrt{\widetilde{c}^{2}-\widetilde{d}^{2} }$ and $\delta^{-1}\equiv\left(\frac{\widetilde{c}-\widetilde{d}}{\widetilde{c}+\widetilde{d}}\right)$.

According to the definition of the generating functions 
(\ref{eq.4.1}),(\ref{eq.4.7}) the instantaneous correlation functions of the dual model (\ref{eq.3.8})
are obtained by taking partial derivatives of  ${\cal V}^{\pm}(\{\eta,j\},t)$ (\ref{eq.4.17}). If one considers the mean-value of an observable $O$ of the original DPAC process (\ref{eq.1.1}), its dual counterpart is  denoted by $\langle \hat{O}(t)\rangle$ as:
$\langle \hat{O}(t)\rangle\equiv \langle \hat{\widetilde{\chi}}|\hat{O} e^{-\hat{H}^{DPAC}t}|\hat{P}(0)\rangle\equiv \langle \widetilde{\chi}|V(V^{-1} O V)(V^{-1}e^{-H^{DPAC}t}V) V^{-1} |P(0)\rangle \equiv \langle O(t)\rangle$, and therefore, the dual counterpart of the observable $O$  (say $n_{j}\equiv \frac{1}{2}(\openone - \sigma_{j}^{z})$) is  $\hat{O}\equiv V^{-1} O V$ ($\hat{n}_{j}\equiv \frac{1}{2}(\openone - \hat{\sigma}_{j-1}^{z}\hat{\sigma}_{j}^{z})$).

For the unbiased DPAC model (i.e. Lushnikov's model, where $h=h'$), we thus recover Glauber's original result \cite{Glauber}. For the biased DPAC situation, the generating function (\ref{eq.4.17}) not only provides the complete solution of the DPAC model but also the complete solution of the {\it biased generalization} of Glauber's model with transition probability $\omega(\sigma_{i}\rightarrow -\sigma_{i})=D\left (1-\frac{d}{2D}\sigma_{i}(\sigma_{i-1}+\frac{c}{d}\sigma_{i+1} )\right)$, where  the $\sigma_{i}=\pm 1$ are the usual spin variables and we assume the rate $D=b/2$. For this model we obtain:$\langle\sigma_{j}(t) \rangle
=\frac{\partial {\cal V}^{-}(\{\eta,j\})}{\partial \eta_{j}}|_{\{\eta\}=0}= e^{-bt}\sum_{k}\langle \sigma_{k}(0)\rangle \delta^{\frac{k-j}{2}}I_{k-j}(\triangle t)$. Relabelling the sites which now run from $-\frac{L}{2}+1$ to $\frac{L}{2}$, the long-time behaviour follows as $\langle \sigma_{j}(t)\rangle -\langle \sigma(0)\rangle e^{-(b-\widetilde{c})t}\sim \frac{e^{-(b-\triangle)t}}{\delta^{-j}\sqrt{\triangle t}}f(\delta)$, where $\delta^{-1}\equiv(\frac{\widetilde{c}-\widetilde{d}}{\widetilde{c}+\widetilde{d}})$, $\triangle\equiv\sqrt{\widetilde{c}^{2}-\widetilde{d}^{2}}$ and $f(u)\equiv \sum_{k}e^{iuk}(\langle \sigma_{k}(0)\rangle-\langle \sigma(0)\rangle )$, which is considered to be an analytic function. This result generalizes recent results \cite{Agham} obtained for the biased {\it zero-temperature} Glauber's model.
\section{ Density and Correlation functions of the DPAC model}
\subsection{ Density and Correlation functions for arbitrary initial states}
In the previous section, we have obtained an explicit expression for the generating function of the dual model (see (\ref{eq.3.6})) and we have shown how to compute correlation functions for the dual model (\ref{eq.3.8}). In this section we show how to relate correlation functions of the dual model (\ref{eq.3.8}) to the correlation function of the original DPAC model (\ref{eq.1.1}).

Here we are especially interested in density-density correlation functions:
$\langle n_{j_{1}} \dots  n_{j_{n}} \rangle(t)=\frac{1}{2^{n}}
\langle (1- \hat{\sigma}_{j_{1}-1}^{z}\hat{\sigma}_{j_{1}}^{z} )(1- \hat{\sigma}_{j_{2}-1}^{z}\hat{\sigma}_{j_{2}}^{z} )\dots (1- \hat{\sigma}_{j_{n}-1}^{z}\hat{\sigma}_{j_{n}}^{z}) \rangle(t), \;\; (j_{1}<j_{2}<\dots<j_{n})$, where the symbol $\hat{\sigma}^{z}$ means that the mean-value $\langle\hat{\sigma}^{z}\rangle(t)$ is taken with respect to the dual model (\ref{eq.3.8}).

In what follows, we  need to know how to connect initial correlation functions of the dual model (\ref{eq.3.8}) with initial those of the original DPAC model (\ref{eq.1.1}):
 $\langle \hat{\sigma}_{j_{1}}^{z}\dots \hat{\sigma}_{j_{2n}}^{z}\rangle (0)
=\left\langle \prod_{j_{1}<j\leq j_{2n}}(1-2n_{j}(0))\right\rangle, 
(j_{1}<j_{2}<\ldots<j_{2n})$

In particular, for the density, we have 
\begin{eqnarray}
\label{eq.5.3}
&&\langle n_{r} \rangle(t)=\frac{1}{2}\left(1-\frac{\partial^{2}{\cal V^{+}}}{\partial \eta_{r-1}\partial_{r}}\right)|_{\{\eta\}=0} =
\frac{1}{2}\left[1- z + e^{-2bt}\sum_{n>0} z^{n}\left\{\
I_{n-1}(2\widetilde{c}t) - I_{n+1}(2\widetilde{c}t)\right\} \right]\nonumber\\
&-&\frac{e^{-2bt}}{2}\sum_{j_{2}>j_{1}}\left\langle\prod_{j_{1}<j\leq j_{2}}(1-2n_{j}(0))\right\rangle \delta^{\frac{j_{1}+j_{2}-2r+1}{2}}\left\{
I_{j_{1}+1-r} (\triangle t)I_{j_{2}-r}(\triangle t ) 
-I_{j_{1}-r}(\triangle t )I_{j_{2}+1-r}(\triangle t) 
\right\},
\end{eqnarray}
where $\triangle\equiv\sqrt{\widetilde{c}^{2}-\widetilde{d}^{2}} $, $\delta^{-1}\equiv \left(\frac{\widetilde{c}-\widetilde{d}}{\widetilde{c}+\widetilde{d}}\right) $ and we have relabelled the sites indices which now run from $-\frac{L}{2}+1$ to $\frac{L}{2}$.

We can also obtain the non-instantaneous two-point correlation functions :
\begin{eqnarray}
\label{eq.5.3.1}
&&\langle n_{r}(t)n_{s}(0)\rangle
=
\frac{1}{2}\left[1- z + e^{-2bt}\sum_{n>0} z^{n}\left\{\
I_{n-1}(2\widetilde{c}t) - I_{n+1}(2\widetilde{c}t)\right\} \right]
\nonumber\\
&-&\frac{e^{-2bt}}{2}\sum_{j_{2}>j_{1}}\left\langle n_{s}(0)\prod_{j_{1}<j\leq j_{2}}(1-2n_{j}(0))\right\rangle \delta^{\frac{j_{1}+j_{2}-2r+1}{2}}\left\{
I_{j_{1}+1-r} (\triangle t)I_{j_{2}-r}(\triangle t ) 
-I_{j_{1}-r}(\triangle t )I_{j_{2}+1-r}(\triangle t) 
\right\}
\end{eqnarray}
To obtain the long-time behaviour of (\ref{eq.5.3.1}), we introduce the Fourier transform of the initial state from which we have substracted the homogeneous part:

 $f(u,v,s)\equiv \sum_{j_{1},j_{2}}e^{i(uj_{1}+vj_{2})}\left(\langle n_{s}(0)\prod_{j_{1}<j\leq j_{2}}(1-2n_{j}(0))\rangle-(1-2\rho(0))^{j_{2}-j_{1}}\right)\Theta(j_{2}-j_{1})$, where $\Theta(j_{2}-j_{1})=1$ if $j_{2}>j_{1}$ and vanish otherwise. With the {\it steepest-descendent method}, we obtain for $|r|,|s|\ll L\rightarrow \infty$ and $bt,|\widetilde{c}|t, \triangle t \gg 1$:
\begin{eqnarray}
\label{eq.5.3.2}
\langle n_{r}(t)n_{s}(0)\rangle -\rho(t)\sim \frac{e^{-2
(b-\triangle)t}}{\delta^{-r}(\triangle t)^{2}}f(\delta,\delta,s),
\end{eqnarray}
where $\rho(t)$ is the translationally invariant and uncorrelated density studied in (\ref{eq.6.10}-\ref{eq.6.19.8}). From (\ref{eq.5.3.2}), we see that the decay of the  connected correlation functions (\ref{eq.5.3.1}) depends on the bias and is in the form of an exponential, except for the unbiased case (i.e. $\widetilde{d}=0$) and when there is no pair-creation (i.e. $b=\widetilde{c}$). In the latter case the decay follows the power-law $\sim (\widetilde{c} t)^{-2}$. For the case $\widetilde{d}<0$, because of the drift, the rightmost sites tend more rapidly to their steady-state. The effect of the initial state appears through the function $f$, which is assumed to be analytic.

All the  the multipoint correlation functions can be obtained in a similar and systematic way.

\subsection{Density for translationally invariant and uncorrelated initial states}
In this subsection we check the  results for  the density
(\ref{eq.5.3})  in the translationally invariant DPAC model against previous results obtained for some particular initial states (lattice with initial density $\rho(0)=0,1/2,1$), directly by  the  {\it free-fermionic approach} \cite{Grynberg1,Santos,Alcaraz,Oliveira,Mobar1}. As original results, we obtain the exact asymptotic behaviour of the density for uncorrelated and translationally invariant states of arbitrary initial density (in the presence as well as in the absence of pair-creation). 

As one can check by direct computation, for tranlationally invariant systems, the density in the DPAC model is independent of the bias $\widetilde{d}$ and thus the density of Lushnikov's model coincides with the density of the generalized DPAC model (for the class of translationally invariant systems under consideration here). It has been shown to be a general property of the  DPAC model, which holds true for every {\it instantaneous} correlation functions \cite{Schutz2}. We have  seen in (\ref{eq.5.3.1},\ref{eq.5.3.2}) that this property is lost for non-instantaneous correlation functions.  

For translationally invariant systems and with help of the properties of Bessel functions, (\ref{eq.5.3}) leads to $\langle\hat{\sigma}^{z}_{r}\hat{\sigma}^{z}_{r+s}\rangle(t)=
\langle\hat{\sigma}^{z}_{0}\hat{\sigma}^{z}_{s}\rangle(t)
=z^{s}+e^{-2bt}\sum_{s'>0}\left(
\langle\hat{\sigma}^{z}_{0}\hat{\sigma}^{z}_{s'}\rangle(0)  -z^{s'}\right) \left(I_{s-s'}(2\widetilde{c}t) 
- I_{s+s'}(2\widetilde{c}t)\right)$,  ($s>0$).
 Where, $z\equiv \frac{b-\sqrt{b^{2}-\widetilde{c}^{2}}}{\widetilde{c}}$.
This result coincides with Glauber's original one \cite{Glauber}.

Let us consider the uncorrelated case where initially $\langle n_{i_{1}} \dots
 n_{i_{n}} \rangle (0)=(\rho(0))^{n}$, ($i_{n}>i_{n-1}>\dots>i_{1}$). This implies:
 $\langle\hat{\sigma}^{z}_{s'}\hat{\sigma}^{z}_{0}\rangle(0)=
\langle (1-2n_{1})\dots (1-2n_{s'}) \rangle(0)=(1-2\rho(0))^{s'}$
and therefore,
\begin{eqnarray}
\label{eq.6.10}
\rho(t)=
\frac{1}{2}\left(1-\langle\hat{\sigma}^{z}_{0}\hat{\sigma}^{z}_{1}\rangle(t) \right)
&=&\frac{(1-z)}{2}-\frac{e^{-2bt}}{2}\sum_{s'>0}\left( (1-2\rho(0))^{s'}   -z^{s'}\right) \left(I_{s'-1}(2\widetilde{c}t) - I_{s'+1}(2\widetilde{c}t)\right)
\end{eqnarray}

To study expression (\ref{eq.6.10}) we notice that it can be rewritten as
\begin{eqnarray}
\label{eq.6.19.1}
\rho(t)-\rho(\infty)=\frac{e^{-2bt}sign \widetilde{c}}{2} \left\{\left[(|z|-y)I_{0}(2|\widetilde{c}|t)+ 
(z^{2}-y^{2})I_{1}(2|\widetilde{c}|t)\right]
+\sum_{n\geq 2}\left((1-y^{2})y^{n-1} -
(1-z^{2})|z|^{n-1}\right)I_{n}(2|\widetilde{c}|t)\right\}
\end{eqnarray}
where $y\equiv (1-2\rho(0))sign\widetilde{c}$.

For the case where $\widetilde{c}=0$ with $\epsilon=\epsilon'>0$ the density (\ref{eq.6.19.1}) reads $\rho(t)=\frac{1}{2}+(\rho(0)-\frac{1}{2})e^{-4\epsilon t}$. From now we focus on the cases where $|\widetilde{c}|\neq 0$.

In the absence of pair-creation (i.e. $\epsilon'=0$, $b=\widetilde{c}=\epsilon>0$), we simply have (imposing $z=1$, in (\ref{eq.6.19.1}))
\begin{eqnarray}
\label{eq.6.19.2}
\rho(t)= \rho(0)e^{-2{\widetilde c}t}\left[I_{0}(2\widetilde{c}t) +\sum_{n\geq 1}(1-2\rho(0))^{n-1}I_{n}(2\widetilde{c}t)\right],
\end{eqnarray}
this result coincides with the result obtained in \cite{Henkel}.

Taking into account the asymptotic behavior  of the Bessel functions $I_{n}(x)$ and collecting terms,  we arrive at the  asymptotic behaviour of the density.

In the massive (non-universal regime) when $bt>|c|t \gg 1, u=L^{2}/|c|t \ll 1$ and $|y|, z<1\Rightarrow
0<\rho(0)<1$, we have:
\begin{eqnarray}
\label{eq.6.19.3}
\rho(t)-\rho(\infty)\sim\frac{ e^{-2(b-|\widetilde{c}|)t} sign \widetilde{c}}{32 |\widetilde{c}|t \sqrt{\pi|\widetilde{c}|t}}
\left[(|z|-y)(1-3(y+|z|)) + \frac{|z|(|z|+1)(15-8|z|-3z^2)}{(1-|z|)^{2}}
-\frac{y(y+1)(15-8y-3y^2)}{(1-y)^{2}}  \right] 
\end{eqnarray}

The validity of (\ref{eq.6.19.3}) is restricted to $|y|<1$, i.e. to $0<\rho(0)<1$, which corresponds to the convergence domain of the geometrical series occuring in computing (\ref{eq.6.19.3}). The cases $\rho(0)=0$ and  $\rho(0)=1$ 
correspond to $y=1$ (for $\rho(0)=0)$ and $y=-1$ (for $\rho(0)=1)$. As $|z|<1$, the residual summation over the expansion of $z^{n-1}I_{n}(2|\widetilde{c}|t)$ can be carried out, leading to  
\begin{eqnarray}
\label{eq.6.16}
\rho(t)-\rho(\infty)
\sim\left\{
\begin{array}{l l l l}
-\frac{e^{-2(b-|\widetilde{c}|)t}}{\sqrt{\pi |\widetilde{c}|t}} &\mbox{, if $\rho(0)=0$ and $\widetilde{c}>0$  }\\
-\frac{e^{-2(b-|\widetilde{c}|)t}}{32(\pi |\widetilde{c}|t)^{3/2}}\left(
1-3(|z|-1)+\frac{|z|(1+|z|)}{1-|z|}(15-8|z|-3z^{2})\right) &\mbox{, if $\rho(0)=0$ and $\widetilde{c}<0$  }\\
\frac{e^{-2(b-|\widetilde{c}|)t}}{32(\pi |\widetilde{c}|t)^{3/2}}\left(1-3(|z|-1)+\frac{|z|(1+|z|)}{1-|z|}(15-8|z|-3z^{2})\right) &\mbox{, if $\rho(0)=1$ and $\widetilde{c}>0$  }\\
\frac{e^{-2(b-|\widetilde{c}|)t}}{\sqrt{\pi |\widetilde{c}|t}} &\mbox{, if $\rho(0)=1$ and $\widetilde{c}<0$  }
\end{array}
\right.
\end{eqnarray}
Results (\ref{eq.6.16}) coincide with results obtained in \cite{Grynberg1,Mobar1}.

In the {\it critical case}, when there is no pair-creation ($\epsilon'=0, b=\widetilde{c}>0,z=1$), we have for  $bt=\widetilde{c}t \gg 1, u=L^{2}/|\widetilde{c}|t \ll 1$ and $|y|<1 \Rightarrow \rho(0)<1$.

\begin{eqnarray}
\label{eq.6.19.4}
\rho(t)&=&\frac{1}{2\sqrt{\pi \widetilde{c}t}}\left(1-\frac{1}{16 \widetilde{c} t}\left[
(1-y)(3y +2)-y(1+y)\frac{15-8y-3y^{2}}{(1-y)^{2}}\right]\right) +{\cal O}\left((\widetilde{c}t)^{-5/2} \right)
\end{eqnarray}
This result is restricted to $0<\rho(0)<1$. When initially there are no particles, $\rho(0)=0, y=1$, no dynamics take place. The case of an initially full lattice ($y=-1, \rho(0)=1$) yields: $\rho(t)_{|_{\rho(0)=1}}=e^{-2\widetilde{c}t}I_{0}(2\widetilde{c}t)=\frac{1}{2\sqrt{\pi \widetilde{c}t}}(1+{\cal O}((\widetilde{c}t)^{-1}))$. 

A similar asymptotic results would have been obtained for the case where there are only {\it pairs created} and no annihilation ($b=\widetilde{c}=\epsilon'>0$ and $\epsilon=0$). In this case $z=-1$ and $\rho(\infty)=1$. With (\ref{eq.6.19.1}), we obtain in the asymptotic regime ($\epsilon't\gg 1$, for an initially partially filled lattice, $0\leq\rho(0)<1$) a critical decay of the density: $\rho(t)=1-\frac{1}{2\sqrt{\pi bt}}\left(1-\frac{1}{16 bt}\left[
(1-y)(3y +2)-y(1+y)\frac{15-8y-3y^{2}}{(1-y)^{2}}\right]
\right)+ {\cal O}((bt)^{-5/2})$.

So far we have considered asymptotic behaviour for times which were much larger than the typical times of diffusion (we had $u=2L^{2}/|\widetilde{c}|t \ll 1$). Now,  we study the asymptotic behaviour of the density, both in the massive and critical regimes, for typical times of order of the diffusion time, i.e. $u=2L^{2}/|\widetilde{c}|t \sim 1$.

In the massive, non-universal regime, when $bt>\widetilde{c}t \gg 1, u=2L^{2}/|\widetilde{c}|t \sim 1$, from (\ref{eq.6.19.1}), we obtain ($\forall y$, i.e. $\forall  0\leq\rho(0)\leq 1$):
\begin{eqnarray}
\label{eq.6.19.5}
\rho(t)-\rho(\infty)&=&\frac{e^{-2(b-|\widetilde{c}|)t}}{4\sqrt{\pi|\widetilde{c}| t}} \left[|z|(|z|+1)-y(y+1)+\sum_{n\geq 2} \left(\{(1-y^{2})y^{n-1} 
-(1-z^{2})|z|^{n-1}\}e^{-\frac{n^{2}}{4|\widetilde{c}|t}}\right) 
+{\cal O}\left((|\widetilde{c}|t)^{-1}\right)
\right]
\end{eqnarray}

In the critical regime  when $\epsilon' =0$ and $bt=|\widetilde{c}|t \gg 1$, with $ u=L^{2}/|\widetilde{c}|t \sim 1$, from (\ref{eq.6.19.1}), we obtain ($0<\rho(0)\leq 1$):
\begin{eqnarray}
\label{eq.6.19.6}
\rho(t)&=&\frac{1}{2\sqrt{\pi\widetilde{c}t}}
\left[\left(1-\frac{y(y+1)}{2}\right)+\frac{1-y^{2}}{2}\sum_{n\geq 2} y^{n-1}e^{-\frac{n^{2}}{4\widetilde{c} t}} \right] + {\cal O}\left(\frac{1}{(\widetilde{c}t)^{-3/2}} \right)
\end{eqnarray}

If one considers $\frac{n^{2}}{4\pi|\widetilde{c}|t}<\frac{L^{2}}{4\pi|\widetilde{c}|t}=u/8\pi \ll 1$, we recover the universal regime (\ref{eq.6.19.4}). A similar study can be performed for the case where there is no pair-annihilation, but only pair-creation ($\epsilon=0$, $b=\widetilde{c}=\epsilon'>0$, for a partially filled initial lattice: $\rho(0)<1$) and, we  have $\rho(t)\sim 1-\frac{1}{2\sqrt{\pi\widetilde{c}t}}
\left[\left(1-\frac{y(y+1)}{2}\right)+\frac{1-y^{2}}{2}\sum_{n\geq 2} y^{n-1}e^{-\frac{n^{2}}{4 \widetilde{c} t}} \right] $.

Let us conclude this subsection with the study of the so-called {\it staggered current} for the (bias-)DPAC model. This quantity has been introduced by Grynberg et al. \cite{Grynberg1} to measure the flux of particles due to the bias (so it vanishes for Lushnikov's model): it is defined as ${\cal J}(t)\equiv
\langle h'n_{m}(1-n_{m+1}) - h n_{m+1}(1-n_{m}) \rangle(t) $. In \cite{Grynberg1} this quantity was computed for an  initially empty lattice. Here we  obtain the  exact expression of this quantity for translationally invariant (and uncorrelated, but random) initial states with arbitrary initial density $\rho(0)$.

For a translational invariant system, the expression of the {\it staggered current} is ${\cal J}(t)=-\widetilde{d}\left(\rho(t) -\langle n_{m+1} n_{m} \rangle (t)\right)=-\frac{\widetilde{d}}{4}\left(1-\langle \hat{\sigma}_{j}^{z} 
\hat{\sigma}_{j+2}^{z}\rangle\right)$. With help of results of this section, we then find:
\begin{eqnarray}
\label{eq.6.19.8}
{\cal J}(t)=\frac{\widetilde{d}}{4}\left\{\rho(\infty)(\rho(\infty)-1)+e^{-2bt}
\sum_{s'>0}\left[(1-2\rho(0))^{s'}-z^{s'}\right]\left(I_{s'-2}(2\widetilde{c}t)-I_{s'+2}(2\widetilde{c}t) \right)\right\}
\end{eqnarray}
The long-time behavior of this quantity follows similarly as in (\ref{eq.6.19.3}-\ref{eq.6.19.6}).

\subsection{Density for translationally invariant but correlated initial states}
In the previous  subsection, we have obtained exact  results for the ordered DPAC model for a class of translationally invariant and uncorrelated initial state. It is also instructive to consider  {\it translationally invariant but correlated} initial states of the form ${\cal F}(\{\sigma_{j}^{x}\sigma_{j+1}^{x}\})|0\rangle$, which according to  (\ref{eq.3.6}) are transformed into $V^{-1}{\cal F}(\{\sigma_{j}^{x}\sigma_{j+1}^{x}\})V |0\rangle={\cal F}(\{\sigma_{j}^{x}\})|0\rangle $, where ${\cal F}(O)$ is a functional of the operator $O$.
 For translationally invariant systems the density is independent of the bias \cite{Schutz2}. In the sequel, we focus on two classes of  translationally invariant but correlated initial states of the form   ${\cal F}(\{\sigma_{j}^{x}\sigma_{j+1}^{x}\})|0\rangle$, namely:

i)
\begin{eqnarray}
\label{eq.6.20}
|P(0)\rangle \equiv \prod_{j}\left(\frac{1+\mu}{2}+ \frac{1-\mu}{2} \sigma_{j}^{x}\sigma_{j+1}^{x}\right)|0\rangle
\end{eqnarray}
We have the initial correlations:
$\langle \sigma_{j}^{z}\rangle (0) =\mu$, $\langle\sigma_{j}^{z} \sigma_{j'\neq j}^{z}\rangle (0) = \mu^{2}$ and so the states (\ref{eq.6.20}) correspond to uncorrelated states for the {\it dual model}. However, via the duality transformation  (\ref{eq.3.6}), this state is related to a correlated initial state in the original DPAC model (\ref{eq.1.1}) \cite{Schutzrev,Santos0}. 
For the density of the DPAC model, with initial states (\ref{eq.6.20}), we have (see (\ref{eq.5.3}))
\begin{eqnarray}
\label{eq.6.21}
\rho(t)=\frac{1-z}{2}+
 \frac{e^{-2bt}}{2}\sum_{s>0}z^{s}\left(I_{s-1}(2\widetilde{c}t) -I_{s+1}(2\widetilde{c}t) \right) -\frac{\mu^{2}e^{-2bt}}{2}\left( I_{0}(2\widetilde{c}t)+ I_{1}(2\widetilde{c}t)  \right),
\end{eqnarray}

When ${\widetilde c}=0$, with $\epsilon=\epsilon'>0$, this expression reduces to $\rho(t)=\frac{1-\mu^{2}e^{-4\epsilon t} }{2}$. Hereafter, we focus on the more interesting case $\widetilde{c}\neq 0$. 

In the first term of the r.h.s of (\ref{eq.6.21}), we recognize the expression of $\rho(t)-\rho(\infty)$ for an initially uncorrelated state with  $\rho(0)=1/2$ (\ref{eq.6.10}). When there is no creation of pairs of particles (i.e. $\epsilon'=0 \Rightarrow b=\widetilde{c}=\epsilon>0, z=1$),
\begin{eqnarray}
\label{eq.6.22}
\rho(t)&=&
\frac{e^{-2\widetilde{c}t} (1-\mu^{2})}{2}\left[I_{0}(2\widetilde{c}t) + I_{1}(2\widetilde{c}t)\right]
\end{eqnarray}
 In fact, such initial ({\it correlated}) states have been considered previously  by Santos \cite{Santos0} who computed the density for the diffusion-annihilation version of the DPAC-model (i.e. $\epsilon'= 0 \Rightarrow z= 1$) using Jordan-Wigner transformation and the {\it free-fermionic procedures}. 

Let us study the asymptotic behaviour of the density for this (correlated) initial state (\ref{eq.6.20}). To this end, we proceed as in the previous subsection: we begin with the massive, non-universal regime, where $bt>|c|t \gg 1$, with $u=L^{2}/|c|t \ll 1$.
In this regime, the density decays as $\rho(t)-\rho(\infty)=
-\frac{e^{-2(b-\widetilde{c})t}}{4\sqrt{\pi|\widetilde{c}|t}}\left(\mu^{2}(1+sign\widetilde{c})+{\cal O}((|\widetilde{c}|t)^{-1})\right)$. On the other hand, in the {\it critical regime} (when $\epsilon'=0, b=\widetilde{c}=\epsilon>0, \widetilde{c}t\gg 1$, with $u=2L^{2}/|\widetilde{c}|t \ll 1$), we have the following power-law decay: $\rho(t)\sim
\frac{1-\mu^{2}}{2\sqrt{\pi \widetilde{c} t}}\left(1-\frac{1}{8\widetilde{c}t}\right)$

As noticed in \cite{Santos0}, we see that although the initial state is correlated, the long-time behaviour of the density decays as  $t^{-\frac{1}{2}}$, as in the uncorrelated cases. The interesting point for the DPA model (where $\epsilon' =0$) is however that the dynamics, though {\it critical}, is no longer universal: the amplitude of the density (i.e. the term $1-\mu^{2}$) depends on the initial state. We infer that in this case initial correlations don't  renormalize the dynamical exponent, yet affect the amplitude in a non-universal manner.

Similarly, a power-law decay is obtained in the case where there is no pair-annihilation (i.e. $b=-\widetilde{c}=\epsilon'>0$):
$\rho(t)\sim 1-\frac{1}{2\sqrt{\pi b t}}$. Notice however that the asymptotic behavior is independent of $\mu$ and so is {\it universal}.

ii) 
Let us now consider another class of translationally-invariant, but {\it correlated}, initial states, namely ($\mu>0$):
\begin{eqnarray}
\label{eq.6.23}
|P(0)\rangle \equiv
\frac{1}{\alpha} \prod_{j}\left(\frac{1+\mu}{2}+ \frac{1-\mu}{2} \sigma_{j}^{x}\sigma_{j+2}^{x}\right)V|0\rangle
\end{eqnarray}
%,
for which initial correlation functions are: $\langle\sigma_{s\geq 0}^{z}\sigma_{0}^{z} \rangle(0)=e^{-\beta}\sinh\beta \left(1+\frac{\delta_{s,1}}{2}+(1-\delta_{s,0}-2\delta_{s,1}) e^{-\beta}\sinh\beta \right) $ and $\langle \sigma_{j}^{z}\rangle(0) =
e^{-\beta}\sinh\beta $, where 
$\beta\equiv\ln\frac{1}{\sqrt{\mu}}$. The initial states are therefore correlated both  for the dual model and  for the DPAC model.

Notice that here the operators coding correlations act are {\it not} nearest-neighbours
preventing a direct Jordan-Wigner transformation of the expression (\ref{eq.6.23}), i.e.  a direct
{\it free-fermion approach}.
 
The computation of the density yields:
\begin{eqnarray}
\label{eq.6.24}
\rho(t)&=&
\frac{1-z}{2}+\frac{e^{-2bt}}{2}\sum_{s>0} z^{s} \left(I_{s-1}(2\widetilde{c}t)-I_{s+1}(2\widetilde{c}t)\right)-\frac{e^{-2bt-\beta}}{2}\sinh\beta (1+e^{-\beta}\sinh\beta)
\left(I_{0}(2\widetilde{c}t)+I_{1}(2\widetilde{c}t)\right) \nonumber\\
&+&\frac{\sinh\beta e^{-\beta}}{2}\left(2 e^{-\beta}\sinh\beta -1\right)
e^{-2bt}\left(I_{0}(2\widetilde{c}t)-I_{2}(2\widetilde{c}t)  \right)
\end{eqnarray}

In the first term of the r.h.s. of (\ref{eq.6.24}),  we recognize the expression for the uncorrelated density $\rho(t)-\rho(\infty)$ with  an uncorrelated initial state with $\rho(0)=1/2$ (\ref{eq.6.10}).

When $\widetilde{c}=0$ with $\epsilon=\epsilon'>0$, (\ref{eq.6.24}) reduces to the following expression:
$\rho(t)=\frac{1}{2}-e^{-\beta}\sinh \beta (\frac{3}{2}-e^{\beta}\sinh\beta)e^{-4\epsilon t}  $. Below, we focus on the case where $\widetilde{c}\neq 0$.

When there are no pairs created (i.e. $\epsilon'\equiv 0, b=\widetilde{c}=\epsilon>0$ and thus $z=1$), this expression reduces to:
\begin{eqnarray}
\label{eq.6.25}
\rho(t)=\left[1-e^{\beta}\sinh\beta (1+e^{-\beta}\sinh\beta)\right]
\frac{e^{-2\widetilde{c}t}}{2}\left(I_{0}(2\widetilde{c}t)+I_{1}(2\widetilde{c}t)\right) +\frac{\sinh\beta e^{-\beta}}{2}\left(2\sinh\beta e^{-\beta}-1\right)
e^{-2\widetilde{c}t}\left(I_{0}(2\widetilde{c}t)-I_{2}(2\widetilde{c}t)  \right)
\end{eqnarray}

Let us now investigate the asymptotic behaviour of the density for this (correlated) initial state (\ref{eq.6.23}). This is done as in the previous subsection. We begin with the massive, non-universal regime, where $bt>|c|t \gg 1$, and ($b>\widetilde{c}$ implies $|z|<1$) with $u=L^{2}/|c|t \ll 1$.
In this regime, the density decays as $\rho(t)-\rho(\infty)=
-\frac{(1+sign \widetilde{c})e^{-2(b-|\widetilde{c}|)t}}{4\sqrt{\pi|\widetilde{c}|t}} \left[e^{-\beta}\sinh \beta (1+e^{-\beta}\sinh \beta)+{\cal O}((|\widetilde{c}|t)^{-1})\right]$.

On the other hand, in the {\it critical regime} (when $\epsilon'=0, b=\widetilde{c}=\epsilon>0, ct\gg 1$ and $z=1$, with $u=2L^{2}/|\widetilde{c}|t \ll 1$), we have a power-law decay: $\rho(t)=
\frac{1-e^{-\beta}\sinh \beta \left(1+e^{-\beta}\sinh \beta\right)}{2\sqrt{\pi\widetilde{c}t}} + {\cal O}((ct)^{-3/2})$

As before, in the case where there would be no pair-annihilation (i.e. $b=-\widetilde{c}=\epsilon'>0$), we have a power-law decay of the density:
$\rho(t)\sim 1-\frac{1}{2\sqrt{\pi b t}}$. However in this case the asymptotic behavior is independent of $\mu$ and so is {\it universal}.

Notice that in the {\it critical regime}, the density decays as a power-law : $\rho(t)\sim t^{-1/2}$, as for uncorrelated initial cases (\ref{eq.6.19.4},\ref{eq.6.19.6}). We also remark that the amplitude of the long-time behaviour of the density is non-universal and depends on the initial state through the parameter $\mu$  (with $\beta=\ln \frac{1}{\sqrt{\mu}}$).

We conclude that for the  correlated states under consideration  (\ref{eq.6.20},\ref{eq.6.23}), initial correlations affect the dynamics through the {\it amplitude of the density}, which in absence of pair-creation is no longer universal. We have observed that when there are  pairs created, without pair-annihilation, the density decays, for the two classes of correlated states (\ref{eq.6.20},\ref{eq.6.23}), as  an {\it universal} power-law.
\subsection{Comparison with traditional free-fermion methods}
In this subsection we discuss and compare the approach devised in this work for the study of the DPAC model with the {\it free-fermion methods} \cite{Schutzrev,Grynberg1,Santos0,Lushnikov,Santos,Alcaraz,Schutz2,Mobar1}.

In most {\it traditional free-fermionic methods}, the stochastic Hamiltonian (\ref{eq.1.1}) is recasted, via a Jordan-Wigner transformation, in (free-)fermion form. The resulting quadratic (but generally neither hermitian nor normal) Hamiltonian is then reformulated in the Fourier space. As the resulting Hamiltonian is quadratic, the time dependence of the operators is simple. Such an approach is well adapted for translationally invariant systems, and  has been extensively used for the study of the DPA model (where $\epsilon'=0$) with homogeneous initial density. Explicit results have in particular been obtained for initially empty and full lattice as well as for initial density $\rho(0)=1/2$ (see e.g. \cite{Schutzrev,Santos0,Santos,Schutz2}). 

Another approach consists in diagonalizing the (free-)fermion version of the DPAC {\it stochastic Hamiltonian} through a Bogoliubov-type rotation, dealing, in Fourier space, with a diagonal quadratic form of the so called {\it pseudo-fermions} \cite{Grynberg1,Mobar1}. This method is particularly efficient for the DPAC model in the presence of pair-creation ($\epsilon'>0$), but concrete results only have been obtained for lattice initially empty or completely filled \cite{Grynberg1,Mobar1}.

Let us also mention that some exact results have also been obtained for particular non-uniform initial distribution for the DPA  and DPAC models \cite{Santos,Oliveira}.

Here, to illustrate the  approach devised in this work, in the translationally invariant situation,  we have recovered known results for the density for both DPA and DPAC models (\ref{eq.6.19.2},\ref{eq.6.16},\ref{eq.6.19.4},
\ref{eq.6.19.6},\ref{eq.6.22}) and have extended these results to the case of arbitrary initial density (\ref{eq.6.19.1},\ref{eq.6.19.3},\ref{eq.6.19.5},\ref{eq.6.19.8},\ref{eq.6.21}). These results should, in principle, be also accessible by the {\it traditional methods} described above, but in the case of arbitrary initial density $0\leq \rho(0)\leq 1$, the computations required  are tedious. On the contrary, with the approach devised here, previous results are recovered in a simple and systematic manner. Although the long-time behaviour of the DPA does not depend on $\rho(0)$, this not the case for the general DPAC in the presence of pair-creation, where the dynamics is no longer {\it universal} but  depends on $\rho(0)$ (through the amplitude and the power-law, see (\ref{eq.6.19.1},\ref{eq.6.19.3},\ref{eq.6.16},\ref{eq.6.19.5})). Therefore, when $\epsilon>0, \epsilon'>0$, the study of the long-time behaviour of the density with respect of the initial density is relevant. An important advantage of the present method is the fact that it is based on an explicit generating function which allow the systematic computation of multi-point correlation functions. In addition, the method employed in this work provides explicit results which are beyond {\it traditional methods}: the results (\ref{eq.5.3},\ref{eq.5.3.1},\ref{eq.5.3.2}) are specific to this approach and are valid for arbitrary initial distribution (uniform, non-uniform, correlated, ...). Furthermore, this approach does not make use of the Jordan-Wigner transformation, and so we are not limited to deal with nearest-neighbour correlations (see (\ref{eq.6.23}-\ref{eq.6.25})). Another advantage of the present approach is that it can be extended to the case of site-dependent DPAC model and will also be useful in the study of disordered version of this model \cite{mobar}. In this sense the present approach is {\it complementary} to previous ones.
\section{Density and non-instantaneous correlation functions for DPACI model}

In this section, we consider  the model described by the {\it stochastic Hamiltonian} $H^{DPACI}=H^{DPAC}+H^{input}$ (according to (\ref{eq.1.2})) where $H^{DPAC}$ and $H^{input}$ have been defined previously (in (\ref{eq.1.1}) and (\ref{eq.1.3}), respectively).
 From a theoretical point of view this model is interesting  because it is one case where (non-trivial) nonequilibrium (the detailed balance is violated) steady-states can be computed exactly.

To proceed with  this study, following the original approach of Racz \cite{Racz}, we first establish the equations of motion of the two-point correlation function of the dual counterpart of  these models, i.e. $\hat{H}^{DPAC}$ (defined in (\ref{eq.3.8})) and  $\hat{H}^{input}$ (defined in (\ref{eq.3.11})), respectively.

We  obtain the equations of motion  of the correlation functions as $\frac{d}{dt}\langle \hat{\sigma}_{m}^{z}\hat{\sigma}_{m+l}^{z} \rangle =
-\langle[\hat{\sigma}_{m}^{z}\hat{\sigma}_{m+l}^{z}, \hat{H}^{DPAC}+\hat{H}^{input} ]\rangle$
For the translational invariant situation, we adopt the following  notation ($l>0$): $r_{l}(t)\equiv
\langle \hat{\sigma}_{m}^{z}\hat{\sigma}_{m+l}^{z}  \rangle(t) =
\langle \hat{\sigma}_{0}^{z}\hat{\sigma}_{l}^{z}  \rangle (t)$. For the non-translational invariant situation, we define $r_{m+l,m}(t)\equiv
\langle \hat{\sigma}_{m+l}^{z}\hat{\sigma}_{m}^{z}  \rangle(t)$.

For the translational invariant system, the equation of motion of the two-point correlation functions is ($\widetilde{c}\neq 0$):
\begin{eqnarray}
\label{eq.7.3}
\frac{d}{dt}r_{l}=-2(b+\gamma l\Theta(L-l))r_{l}(t)+ \widetilde{c}\left(r_{l+1}(t) +r_{l-1}(t) \right),
\end{eqnarray}
with the {\it boundary} condition $r_{0}=1$, where $\Theta(L-l)=1$ for $l<L$ and $\Theta(L-l)=0$, for $l=L$.

For the non-translationally invariant case, we consider the unbiased version of the DPAC model ((\ref{eq.1.1}) with $h=h'>0$, i.e. Lushnikov's model), and the equation of motion of the correlation functions then reads ($l>0$):
\begin{eqnarray}
\label{eq.7.4}
\frac{d}{dt}r_{m+l,m}=-2\left[b+\gamma (2m\Theta(L-m)+l)  \right]r_{l}(t)+
 c\left(r_{m+l+1,m}(t) +r_{m+l-1,m}(t) + r_{m+l,m}+1(t) + r_{m+l,m-1}(t)\right) ,
\end{eqnarray}
with the {\it boundary condition} $r_{0,0}=1$.

i) We start with the translationally invariant problem (\ref{eq.7.3}).

The steady state $\bar{r}_{l}$ corresponding to this situation obeys the difference-equation:
\begin{eqnarray}
\label{eq.7.5}
2(b+\gamma l\Theta(L-l))\bar{r_{l}}= \widetilde{c}\left(\bar{r}_{l+1} +\bar{r}_{l-1} \right),
\end{eqnarray}
To solve (\ref{eq.7.5}), we make  the Ansatz $\bar{r}_{l}=A J_{l+\alpha}\left(\widetilde{c}/\gamma\right)$ ( where $J_{\nu}(\omega)$ is the Bessel function of first kind) and take into account  the following  property of Bessel functions:   $J_{\nu+1}(\omega)+J_{\nu-1}(\omega)=\frac{2\nu}{\omega}J_{\nu}(\omega) $.

Therefore: $\alpha=\frac{b}{\gamma}-L\delta_{l,L}$. Taking into account the boundary condition $r_{0}=\bar{r}(0)=1$,  the constant $A$ follows as: $A=\left(J_{\frac{b}{\gamma}}(\widetilde{c}/\gamma)\right)^{-1}$. The steady-state of the density for translationally invariant systems described by (\ref{eq.7.5}) reads: $\bar{r}_{l}=\frac{J_{l+\frac{b}{\gamma}-L\delta_{l,L}}(\widetilde{c}/\gamma)}{J_{\frac{b}{\gamma}}(\widetilde{c}/\gamma) }$. To solve the dynamical equation (\ref{eq.7.3}), we seek  a solution in the form \cite{Racz,Glauber}: $r_{l}(t)=\bar{r}_{l}+\sum_{l'}q_{l'-l}e^{-2\lambda_{l'}\widetilde{c}t}$

With this Ansatz, (\ref{eq.7.3}) reduces to the following difference equation: $q_{l'-l+1}+q_{l'-l-1}=\frac{2}{\widetilde{c}}\left[b-\widetilde{c}\lambda_{l'}-\gamma l\Theta(L-l)\right]q_{l'-l},$
which is solved as above, with an Ansatz: $q_{l'-l}=B_{l'} J_{l'-l+\beta}(\widetilde{c}/\gamma)$. We thus find: $\beta= \frac{b-\widetilde{c}\lambda_{l'}}{\gamma} + 2l-l' -L\delta_{l,L}$, where the {\it spectrum} $\lambda_{l'}$ is  determined  by the boundary condition $r_{0}=1$, which implies $q_{l'}=0$, and imposes the condition:
\begin{eqnarray}
\label{eq.7.9}
J_{\left(b-\widetilde{c}\lambda_{l'}\right)/\gamma}\left(\widetilde{c}/\gamma\right)=0
\end{eqnarray}
The constants $B_{l'}$ are fixed by the initial condition: $\sum_{l'}q_{l'-l}=r_{l}(0)-\bar{r}_{l} =\sum_{l'}B_{l'} J_{\left(b-\widetilde{c} \lambda_{l'}\right)/\gamma +l-L\delta_{l,L}}\left(\widetilde{c}/\gamma\right)$

Finally, from the the spectrum of $\{\lambda_{l'}\}$ defined by (\ref{eq.7.9}) and the $B_{l'}$ defined above , the  solution of the equation (\ref{eq.7.3}) follows
\begin{eqnarray}
\label{eq.7.11}
r_{l}(t)=\bar{r}_{l}+ \sum_{l'} B_{l'}e^{-2\lambda_{l'} \widetilde{c}t}
 J_{\left(b-\widetilde{c} \lambda_{l'}\right)/\gamma +l-L\delta_{l,L}}\left(\widetilde{c}/\gamma\right),
\end{eqnarray}
where the {\it inverse of the relaxation time}, say $\lambda^{\ast}$, is determined from the {\it smallest } zero of the  Bessel function (\ref{eq.7.9}) and the steady-state has been obtained above.

With the definition of the duality transformation (\ref{eq.3.6}), as in (\ref{eq.5.3}), we obtain for the density of particles $\rho(t)$:
\begin{eqnarray}
\label{eq.7.12}
\rho(t)=\frac{1-r_{1}(t)}{2}=\rho(\infty)- \sum_{l'} \frac{B_{l'}}{2}e^{-2\lambda_{l'} \widetilde{c}t}
 J_{\left(b-\widetilde{c} \lambda_{l'}\right)/\gamma +1} \left(\widetilde{c}/\gamma\right),
\end{eqnarray}
where the steady-state is $\rho(\infty)=\frac{1-\bar{r}_{1}}{2}= \frac{1}{2} \left(1-\frac{J_{\left(1+b/\gamma\right)} \left(\widetilde{c}/\gamma\right)}{
J_{\left(b/\gamma\right)} \left(\widetilde{c}/\gamma\right)
 }\right)$.
In the absence of pair-creation ($\epsilon'=0\Rightarrow b=\widetilde{c}=\epsilon>0$), we recover the results of  \cite{Racz} (with $b=\widetilde{c}=\epsilon=1$).

Following reference \cite{Racz}, we can now study the relaxation-time in the limit of {\it a weak source} (i.e. when $\gamma \ll 1$) and analyze its interplay with the pair-creation term (described by ($b-|\widetilde{c}|)/2$). For the 
DPAC model, where $\epsilon'>0$ and $\gamma=0$, we have seen in section V that the density (for translationally invariant systems) decays exponentially fast in time as $\sim\exp(-2(b-|\widetilde{c}|)t)$. In \cite{Racz} it has been shown that in absence of pair-creation (i.e. $b=\widetilde{c}=\epsilon>0$) the density of the DPAI model decays as $\sim e^{-\gamma^{2/3}t}$. Here we will study the case where both pair-creation ($b>\widetilde{c}>0$) and the source term ($\gamma\geq 0$) are present.

As we focus on the smallest value $\lambda^{\ast}$ of the spectrum $\{|\lambda_{l'}|\}$, we consider $\gamma \ll 1$. In this limit,  we obtain an explicit relation which the $\{\lambda_{l'}\}$  have to fulfill in terms of the zero's $a_{l'}$ of Airy function ($Ai(a_{l'})=0$), namely : $\frac{\mu(1+\lambda_{j})-1}{(1-\mu \lambda_{j})^{1/3}}=\left(\frac{\gamma^{2}}{2b^{2}}\right)^{1/3}|a_{j}|$, where $\mu\equiv\frac{\widetilde{c}}{b}$ (with $|\mu|\geq 1$)  and the $a_{l'}$'s are real and negative.

In the presence of pair-creation (i.e. when $b>|\widetilde{c}|$), one has to consider the above equation , which should be solved for the (infinite) set of $a_{j}$'s. In so doing,  we would obtain the spectrum $\{\lambda_{j}\}$,
 which in turn provide the inverse of the relaxation-time $\lambda^{\ast}\equiv min(\{|\lambda_{j}|\} )$. Here, we prefer to focus on  $\gamma \ll 1$, and  $\mu\lesssim 1$, i.e., we assume that the pair-creation term is also ``small'' ($\epsilon' \ll 1$). Therefore, the spectrum obeys the following equation: $\lambda_{j}=\frac{3 \mu}{\mu+2}\left[\left(\frac{\gamma^2}{2b^{2}}\right)^{1/3}|a_{j}|+1-\mu\right]$.
 In the absence of  pair-creation (i.e. $\mu=1$), we recover the result of \cite{Racz}: $\lambda_{j}=\left(\frac{\gamma^2}{2b^{2}}\right)^{1/3}
|a_{j}|$.

The inverse of relaxation time $\lambda^{\ast}$, follows ($\gamma \ll 1, \epsilon' \ll 1$ and $\mu\lesssim 1$)
\begin{eqnarray}
\label{eq.7.14.1}
\lambda^{\ast}=
\frac{3\widetilde{c}}{\widetilde{c}+2b}\left[\left(\frac{\gamma^2}{2b^{2}}\right)^{1/3}
|a_{1}|+\frac{b-\widetilde{c}}{b}\right]
\end{eqnarray}
Thus the long-time behaviour ($bt>|\widetilde{c}|t \gg 1; \gamma \ll 1 \epsilon' \ll 1$ and $\mu\lesssim 1$) of the density reads (see (\ref{eq.7.14.1})) : $\rho(t)-\rho(\infty)\sim -\frac{B^{\ast}}{2}e^{-2\lambda^{\ast}\widetilde{c}t}$,
where $B^{\ast}$ corresponds to the term $B_{l'}$ for which $\lambda_{l'}=\lambda^{\ast}$.

ii) Let us now pass to the non-translationally invariant situation. In this case, we have to solve the linear differential-difference equation (\ref{eq.7.4}). To do this, we follow the same prodedure as above. The steady-state $\bar{r}_{m+l,m}$ is inferred from the Ansatz $\bar{r}_{m+l,m}=
A(J_{m+l+\alpha}\left(c/\gamma\right))^{2}$ and obtain:
$\alpha = \frac{b}{2\gamma}-\frac{l}{2}-L\delta_{m,L}$. Taking into account the boundary condition $\bar{r}_{m,m}=1$  we find ($l>0$): $\bar{r}_{m+l,m}=\left(\frac{J_{m+\frac{l}{2}+\frac{b}{2\gamma} -L\delta_{m,L}}(c/\gamma) }{J_{m+\frac{b}{2\gamma} -L\delta_{m,L}}(c/\gamma)}\right)^{2}$. Hereafter for notational simplicity, we will denote $\lambda_{m',l'}$ simply by $\lambda$.

To solve the dynamical equation (\ref{eq.7.4}), we seek a solution of the form:
$r_{m+l,m}(t)=\bar{r}_{m+l,m}+\sum_{m',l'}e^{-2\lambda ct} q_{m'+l'-m-l,m'-m}$
and obtain the solution of (\ref{eq.7.4}):
\begin{eqnarray}
\label{eq.7.18}
r_{m+l,m}(t)=\bar{r}_{m+l,l}+\sum_{m',l'}A_{m',l'}e^{-2\lambda ct} 
J_{\frac{b-c\lambda}{\gamma}+m-L\delta_{m,L}}(c/\gamma)
J_{\frac{b-c\lambda}{\gamma}+m+l-L\delta_{m,L}}(c/\gamma),
\end{eqnarray}
where the boundary condition $q_{m'+l'-m,m'-m}=0$ determines the spectrum $\{\lambda\}$
through the condition:
\begin{eqnarray}
\label{eq.7.19}
J_{\frac{b-c\lambda}{\gamma}+m-L\delta_{m,L}}(c/\gamma)=0
\end{eqnarray}
The constants $A_{m',l'}$ follow from the initial conditions.

With (\ref{eq.5.3}, \ref{eq.7.18}), we obtain the  density of particles ($m<L$):

\begin{eqnarray}
\label{eq.7.21}
\langle n_{m+1}\rangle (t)= \langle n_{m+1}\rangle (\infty)-\frac{1}{2}\sum_{m',l'}A_{m',l'}e^{-2\lambda ct} 
J_{\frac{b-c\lambda}{\gamma}+m}(c/\gamma)
J_{\frac{b-c\lambda}{\gamma}+m+1}(c/\gamma),
\end{eqnarray}
with
\begin{eqnarray}
\label{eq.7.22}
 \langle n_{m+1}\rangle (\infty)=\frac{1}{2}\left[1-\left(\frac{
J_{m+\frac{1}{2}+\frac{b}{2\gamma}}(c/\gamma)
}{J_{m+\frac{b}{2\gamma}}(c/\gamma)}\right)^{2}\right]
\end{eqnarray}

iii) From the solution (\ref{eq.7.21}) for the density in the non-translationally invariant situation, we can obtain the non-instantaneous two-point correlation functions of the (unbiased) DPACI model.

With $p-1\leq m <L$, $ \langle n_{m+1}(t)n_{p}(0)\rangle- \langle n_{m+1}(\infty)\rangle= 
 -\sum_{m',l'}\frac{B_{m',l'}}{2} e^{-2\lambda ct}J_{\frac{b-c\lambda}{\gamma}+m}(c/\gamma)
J_{\frac{b-c\lambda}{\gamma}+m+1}(c/\gamma)$

The only change from (\ref{eq.7.21}) is in the $B_{m',l'}$ which are fixed from the initial conditions.

The spectrum $\{\lambda\}$  again follows from equation (\ref{eq.7.19})
and  provides the inverse of relaxation time $\lambda^{\ast}$, as
$\lambda^{\ast}=min\left(\{\lambda_{n,j} \} \right) $.

With $n\equiv m-L\delta_{m,L}$, in the limit $\gamma |n|\ll 1$, $\epsilon'\ll 1$, with $\mu \approx 1$, the spectrum satisfy the equation: $\lambda_{n,j}\approx \frac{b+2\gamma n}{c}-\frac{2\gamma}{c(2\mu -1)^{3/2}}
\left[2^{-1/3}|a_{j}|+2\mu n \left(\frac{2\gamma}{b}\right)^{1/3}\right]^{3/2}$, where  the $a_{j}<0$ are again the zeros of the Airy function.

It has to be emphasized that steady-states obtained for this DPACI model are nonequilibrium steady-states, in the sense that they violate the detailed-balance condition. In the DPAC model (where $\gamma=0$), the steady-states were expressed in terms of the quantity $z$ ( with $\rho(\infty)=\frac{1-z}{2}$) first  introduced  by Glauber \cite{Glauber} to describe the  steady-state of the {\it time-dependent Ising model} (with the prescription of recovering, as steady-state, the {\it thermal equilibrium}). Thus, in this sense the  steady-state of the  DPAC model is  reminiscent of the states in  {\it equilibrium statistical mechanics}. On the contrary, the steady-states of the DPACI model are genuine {\it nonequilibrium steady-states}.

\section{Connection with the coagulation model}
In this section we translate the results obtained so far for the DPAC-model to another, free-fermionic, model of stochastic hard-core particles on a lattice, namely the {\it diffusion-limited coagulation} (DC) model. In this model particles can jump from a site $r$ to an adjacent site (provided it was vacant) with rates $h$ ($\emptyset +A \longrightarrow  A+ \emptyset$) and $h'$ ($A+\emptyset \longrightarrow \emptyset +A$). Two adjacent particles at site $r$ and $r-1$ can coagulate with rate  $h$ ($A+A \longrightarrow  A + \emptyset$).  Two adjacent particles at site $r$ and $r-1$ can also coagulate with rate  $h'$ ($A+A \longrightarrow \emptyset +A$):

 It is known that (for initially uncorrelated systems) there is an  exact duality between DPA-model (where there no pair-creation: $\epsilon' \equiv 0$) and the coagulation model (see e.g.\cite{Schutzrev,Santos,Henkel,Krebs} and references therein). Here we briefly recall the transformation connecting these two stochastic models and translate the results of the $DPA$-model.

Applying the {\it similarity}-transformation \cite{Schutzrev,Santos,Henkel,Krebs} $\widetilde{H}={\cal B}H{\cal B}^{-1}\; ; \;
{\cal B}\equiv B\otimes B\otimes \dots \otimes B=B^{\otimes L}$,
where
$
B\equiv
\left(
 \begin{array}{c c }
 1 & -1   \\
 0 & 2 \\
 \end{array}\right),
$
the ({\it stochastic}-)Hamiltonian $H$ of the DPA-model becomes the
  ({\it stochastic}-)Hamiltonian ${\widetilde H}$ of the coagulation-model.

 For the coagulation model (with initial state $|P'(0)\rangle$), we can express the correlation functions with help of correlation function of the DPA model. The density of the DC then reads:$\langle n_{j}(t)\rangle^{DC}
_{|P'(0)\rangle}=2 \langle n_{j} (t)\rangle^{DPA}_{{\cal B}^{-1}|P'(0)\rangle}$. For the two-point correlations functions, we have:$\langle n_{j+r}(t) n_{j}(t) \rangle^{DC}_{|P'(0)\rangle}= 4 \langle n_{j+r}(t) n_{j} (t)\rangle^{DPA}_{{\cal B}^{-1} |P'(0)\rangle}$ and $\langle n_{j+r}(t) n_{j}(0)\rangle^{DC}_{|P'(0)\rangle} =2 \langle n_{j+r}(t) n_{j} (0)\rangle^{DPA}_{ {\cal B}^{-1}|P'(0)\rangle}$

As an illustration, let us now provide, the density of the $DC$ model from the expressions obtained for the DPA-model (where $\epsilon' \equiv 0$) as
(see (\ref{eq.5.3})). The density of the translationally invariant DC model with  uncorrelated initial state $|P'(0)\rangle =(1-\rho(0)\; \rho(0))^{T}$ (of initial density $\rho$), is related to the unbiased version of the DPA model   with initial density $\frac{\rho(0)}{2}$. Thus, in this situation,  with (\ref{eq.5.3}) , we obtain for the density of DC model:
\begin{eqnarray}
\label{eq.8.9}
\rho^{DC}(t)&=&
e^{-2(h+h')t}\left\{I_{0}(2(h+h')t)+I_{1}(2(h+h')t) -\sum_{s>0}(1-\rho(0))^{s}\left(
I_{s-1}(2(h+h')t)-I_{s+1}(2(h+h')t)\right)\right\}
\end{eqnarray}
As for the DPA-model, for translationally invariant systems, the instantaneous
correlation functions {\it do not depend on the  bias} $h-h'$.  We could also extend in a similar manner  all the  results obtained in the previous sections V to the DC model. The discussion of the long-time behaviour of $\rho^{DC}(t)$ follows  directly from that of the DPA model given  in subsection V.B.
\section{Summary and Conclusion}
This work provides a complete and unified solution of some related  one-dimensional nonequilibrium models.

In a  first part we have solved the {\it diffusion-limited pair-annihilation-creation} (DPAC) model (under the free-fermionic constraint) by a novel approach, {\it alternative  and complementary} to the traditional {\it free-fermion methods}. Combining domain-wall duality \cite{Schutzrev,Santos0}  with a recently developped technique \cite{Aliev}, we were able to compute the generating function of the dual of the DPAC model (section IV), i.e.  to obtain correlation functions (section V).
In this paper, we take advantage of this approach to recover results previously obtained (see (\ref{eq.6.19.2},\ref{eq.6.16},\ref{eq.6.19.4},\ref{eq.6.22})) by {\it traditional methods}, extend these (see(\ref{eq.6.19.1},\ref{eq.6.19.3},\ref{eq.6.19.5},\ref{eq.6.19.6},\ref{eq.6.19.8},\ref{eq.6.21})) and derive original ones (see(\ref{eq.5.3},\ref{eq.5.3.1},\ref{eq.5.3.2},\ref{eq.6.24},\ref{eq.6.25})), in a simple manner. The method is very promising for site-dependent and disordered systems \cite{mobar}. We focused on the density, the {\it staggered-current}, and the two-point correlation functions (instantaneous and non-instantaneous) (see (\ref{eq.5.3},\ref{eq.5.3.1},\ref{eq.5.3.2})). In particular we studied explicitly the density (in absence as well as in presence of pair-creation) for translationally invariant systems with arbitrary initial density and have discussed the asymptotic behaviour of these quantities in different regimes.
We obtained explicit expression of the {\it staggered-current} in arbitrary translationally-invariant (uncorrelated but random) initial conditions (see (\ref{eq.6.19.8})).  We investigated the densities for two classes of {\it correlated} (but tranlationally invariant) systems. 
 We have computed explicitly the non-instantaneous two-point correlation functions for initially translationally invariant and uncorrelated (but random) states: these quantities depend on an  eventual bias  (contrary to instantaneous correlation functions). As a by-product we provided the solution of a biased generalization of Glauber's model, which is the dual counterpart of the DPAC model.

In a second part we studied the density and non-instantaneous two-point 
correlation functions of the diffusion-limited pair-annihilation-creation 
model in the presence of source of particles  (DPACI model). Via the domain-wall duality transformation, we solved exactly the equations of motion of the correlation functions of the dual model. We obtained the exact density and steady-states for translationally invariant as well as non translationally-invariant systems. We also computed the non-instantaneous two-point correlation functions.
In particular for the translationally invariant situation in the absence of the pair-creation term, we recovered, the density first computed by Racz \cite{Racz}. We extended
 these results to the case where  both pair-creation term ($\epsilon'>0$) and source term were present ($\gamma>0$) and obtained the relaxation-time for the case when both $\gamma \ll 1$ and $\epsilon' \ll 1$ (see (\ref{eq.7.14.1})).

In the last part of this work, we have shown how to translate the results obtained for the DPA model to the {\it diffusion-limited coagulation} (DC) model.

\section{Acknowledgments}
The support of the Swiss National Fonds is gratefully acknowledged.

%
%%%%%%%%%%%%%%%%%%%%%%%%%%%%%%%%%%%%%%%%%%%%%%%%%%%%%%%%%%%%%%%%%%%%%%%%%%%%%%

%
%
\end{document}